# On the Capacity and Design of Limited Feedback Multiuser MIMO Uplinks [*]

Il Han Kim and David J. Love

School of Electrical and Computer Engineering

Purdue University

EE Building

465 Northwestern Ave.

West Lafayette, IN 47907

(Ph) 765-496-6797

(Fax) 765-494-0880

kim385@ecn.purdue.edu, djlove@ecn.purdue.edu

October 25, 2018

## Abstract

The theory of multiple-input multiple-output (MIMO) technology has been well-developed to increase fading channel capacity over single-input single-output (SISO) systems. This capacity gain can often be leveraged by utilizing channel state information at the transmitter and the receiver. Users make use of this channel state information for transmit signal adaptation. In this correspondence, we derive the capacity region for the MIMO multiple access channel (MIMO MAC) when partial channel state information is available at the transmitters, where we assume a synchronous MIMO multiuser uplink. The partial channel state information feedback has a cardinality constraint and is fed back from the basestation to the users using a limited rate feedback channel. Using this feedback information, we propose a finite codebook design method to maximize sum-rate. In this correspondence, the codebook is a set of transmit signal covariance matrices. We also derive the capacity region and codebook design methods in the case that the covariance matrix is rank-one (*i.e.*, beamforming). This is motivated by the fact that beamforming is optimal in certain conditions. The simulation results show that when the number of feedback bits increases, the capacity also increases. Even with a small number of feedback bits, the performance of the proposed system is close to an optimal solution with the full feedback.

*Index Terms*-Limited feedback, MAC systems, MIMO systems, CSI at receiver (CSIR), CSI at transmitter (CSIT), Rayleigh channels, Grassmannian subspace packing

[*]This work was supported in part by the NSF under grant CCF0513916, the Korea Research Foundation Grant funded by the Korea government (MOEHRD) (KRF-2005-215-D00207), and the SBC Foundation

# 1 Introduction

Multiple access systems are being widely studied to effectively support multiple users at the basestation during uplink transmission. In multiple-input multiple-output multiple access channels (MIMO MACs), it is known that the channel state information at the transmitters (CSIT) helps increase system performance (e.g., sum-rate capacity) regardless of whether it is perfect or partial [1–5]. CSIT can be full channel information [1–3], channel covariance information [4,5], channel mean information [4, 5], etc. Especially in frequency division duplexing (FDD) systems, because channel state information (CSI) is not reciprocal between uplink channel and downlink channel, this information should be fed back from the receiver to the transmitters. In real communication systems, however, the feedback resources allotted for CSI feedback must be a small portion of the data being transmitted on the downlink because feedback bits are overhead that causes performance degradation.

In this correspondence, we focus on limited feedback schemes, where the receiver is assumed to have perfect knowledge of the CSI and the receiver transmits a limited number of feedback bits to the transmitter according to the CSI. It is of our interest to optimize signal covariance matrices of each user using a limited number of feedback bits. The basestation and users are assumed to have knowledge of a codebook of covariance matrices designed offline. The limited number of feedback bits allows each user to identify the covariance matrix chosen by the basestation for them to use during transmission.

The main contributions of this correspondence are given below:

1) In MIMO systems, we prove the coding theorem for MACs when a limited number of feedback bits is available at the transmitters. Based on the proved coding theorem, we give a method to design a codebook (set of covariance matrices) for each user under a finite feedback cardinality constraint to maximize the sum-rate capacity. Assuming $B$ bits of feedback, the basestation selects one codeword that maximizes the sum-rate among $2^B$ elements in a coodebook. This codebook is assumed to be known to both the basestation and the users. The index corresponding to the chosen codebook matrix is fed back to the users.

2) In single user MIMO scenarios, beamforming is optimal under certain conditions [6–10] when feedback is channel mean or channel covariance information. In multiuser scenarios, beamforming is generally not universally optimal, *i.e.*, the full capacity rate region cannot be achieved. However, if some criteria are satisfied, the sum-rate is asymptotically achieved by beamforming. Studied in [11], the optimality of beamforming was derived when channel feedback is perfect. In [4, 5], the optimality of beamforming was proved when feedback is channel covariance or channel mean information. Motivated by these works, we constrain the rank of each user's covariance matrix to one, *i.e.*, each user performs beamforming on his/her own array. We derive the capacity region of the beamforming system and propose two methods to design the beamforming vectors: eigenbeamforming with a codebook designed using Lloyd's algorithm and Grassmannian subspace packing with random power allocation of each user.

## 2 System Overview

In this section, we introduce the system model under consideration. A typical MIMO MAC communication system is presented in Fig. 1. Let $M_t$ be the number of transmit antennas at each user and $M_r$ be the number of receive antennas. In this setup, the received signal is given by

$$\boldsymbol{y}_n = \sum_{k=1}^{K} \boldsymbol{H}_n^{(k)} \boldsymbol{x}_n^{(k)} + \boldsymbol{z}_n = \begin{bmatrix} \boldsymbol{H}_n^{(1)} & \boldsymbol{H}_n^{(2)} & \cdots & \boldsymbol{H}_n^{(K)} \end{bmatrix} \begin{bmatrix} \boldsymbol{x}_n^{(1)} \\ \boldsymbol{x}_n^{(2)} \\ \vdots \\ \boldsymbol{x}_n^{(K)} \end{bmatrix} + \boldsymbol{z}_n \equiv \boldsymbol{H}_n \boldsymbol{x}_n + \boldsymbol{z}_n, n = 1, \ldots, N, \qquad (1)$$

where $K$ is the number of users and $\boldsymbol{x}_n^{(k)} \in \mathcal{X}^{M_t}$ is the transmitted signal from the $k$-th user with $\mathcal{X}$ denoting the transmit symbol alphabet. Here, $n$ is the time index and $N$ is the block length of each code $\boldsymbol{x}_1^N = \{\boldsymbol{x}_1, \ldots, \boldsymbol{x}_N\}$. $\boldsymbol{H}_n^{(k)}$ is the $M_r \times M_t$ channel matrix from the $k$-th user to the basestation. $\boldsymbol{z}_n$ is the complex Gaussian noise with zero mean and covariance $\sigma^2 \boldsymbol{I}$. Here we assume that data symbols of each user are independent of each other, *i.e.*, $E\left[\mathbf{x}_n^{(k)} \left(\mathbf{x}_n^{(l)}\right)^*\right] = \mathbf{O}_{M_t \times M_t}, \forall k \neq l$, where $\mathbf{O}_{M_t \times M_t}$ is $M_t \times M_t$ all zero matrix. We assume a Rayleigh flat fading environment, where the channel independently changes from transmission to transmission. We also assume that the channel state information $U_n \in \mathcal{U}$ and $V_n \in \mathcal{V}$ are available at the transmitters and the receiver, respectively, where $\mathcal{U}$ is the space of CSIT and $\mathcal{V}$ is the space of channel state information at the receiver (CSIR). CSIT $U_n$ is fed back to the transmitters based on the current CSIR $V_n$. We assume that the channel coefficients are estimated perfectly at the receiver. Since the channel fading is assumed to be independent for each $n$, the transmitted symbol is a function of the feedback CSIT $U_n$. We assume noiseless feedback, *i.e.*, the users know $U_n$ perfectly.

Message $W^{(k)}$ for user $k$, $k = 1, \ldots, K$, is encoded to $\mathbf{x}_n^{(k)}$ for $n = 1, \ldots, N$ depending on the feedback $U_n$. These encoded signals are sent through the channel with channel transition probability $p\left(\boldsymbol{y}_1^N | \boldsymbol{x}_1^N, \boldsymbol{H}_1^N\right)$, where $\boldsymbol{H}_1^N = \{\boldsymbol{H}_1, \ldots, \boldsymbol{H}_N\}$ and $\boldsymbol{y}_1^N = \{\boldsymbol{y}_1, \ldots, \boldsymbol{y}_N\}$. After receiving $\mathbf{y}_1^N$, the receiver decodes $\mathbf{y}_1^N$ to $(\hat{W}^{(1)}, \ldots, \hat{W}^{(K)})$. Assuming the discrete memoryless channel model, the system is characterized by the following channel transition probability

$$p\left(\boldsymbol{y}_1^N | \boldsymbol{x}_1^N, \boldsymbol{H}_1^N\right) = \prod_{n=1}^{N} p\left(\boldsymbol{y}_n | \boldsymbol{x}_n, \boldsymbol{H}_n\right). \qquad (2)$$

This is illustrated in Fig. 2.

# 3 Capacity Results

In this section, we give capacity results when complete CSIR and partial CSIT are available at the receiver and the transmitters, respectively. Partial CSIT is assumed to consist of feedback information from the basestation to the users.

The system model of Fig. 2 can be transformed into an equivalent system without CSIR and CSIT [12–16] by seeing $U_n$ as a channel state and $V_n$ as a channel output. Similarly, in this multiple access system, we first encode each $W^{(k)}$ into the matrix $\mathbf{T}_n^{(k)} \in \mathcal{X}^{M_t \times |\mathcal{U}|}$, $n = 1, \ldots, N$, and select one column of $\mathbf{T}_n^{(k)}$ according to the current feedback $U_n \in \mathcal{U}$ such that $\mathbf{x}_n^{(k)} = \mathbf{T}_n^{(k)}(U_n)$. With the probability distribution function $p(\mathbf{H}_n, u_n, v_n)$, the equivalent channel is characterized by the channel transition probability [15]

$$p(\mathbf{y}_n, v_n | \mathbf{T}_n) = \sum_{u_n, \mathbf{H}_n} p(\mathbf{H}_n, u_n, v_n) p(\mathbf{y}_n | \mathbf{T}_n(U_n), \mathbf{H}_n), \qquad (3)$$

where $\mathbf{T}_n = \left[ \left(\mathbf{T}_n^{(1)}\right)^T \cdots \left(\mathbf{T}_n^{(K)}\right)^T \right]^T$. Fig. 3 presents this equivalent system model.

We assume that the CSIR includes both the CSIT component and the channel coefficient matrix, and the receiver feeds back the $U_n$ component of the CSIR to the transmitters. This means that the receive channel state information is given by $V_n = (\mathbf{H}_n, U_n)$ [16]. We assume the finite rate feedback satisfies $|\mathcal{U}| = 2^B$ with $\mathcal{U} = \{u^{(1)}, \ldots, u^{(2^B)}\}$.

To derive the multiple access capacity region, let $\mathcal{S} = \{k_1, \ldots, k_{|\mathcal{S}|}\} \subseteq \{1, \ldots, K\}$ be a user subset. Each user covariance matrix is given by $\boldsymbol{Q}^{(k)}(U_n) = E\left[\mathbf{x}_n^{(k)} \left(\mathbf{x}_n^{(k)}\right)^* \Big| U_n\right]$ for $n = 1, \ldots, N$, $k = 1, \ldots, K$, with the power constraint $\mathrm{Tr}\left(\boldsymbol{Q}^{(k)}(U)\right) \leq P_k$ for $\forall U \in \mathcal{U}$. This means that we do not allow the average sum transmit power conditioned on some feedback information to exceed $P_k$ for any feedback possibility. In this setup, the capacity region is given by the following theorem.

**Theorem 1** *For a fixed mapping from $U$ to $\left\{\boldsymbol{Q}^{(k)}(U)\right\}_{k=1}^{K}$ such that $\mathrm{Tr}\left(\boldsymbol{Q}^{(k)}(U)\right) \leq P_k$ for $k = 1, \ldots, K$, the capacity region is the $K$-dimensional polyhedron*

$$\left\{ \boldsymbol{R} : \sum_{k \in \mathcal{S}} R^{(k)} \leq E_V \left[ \log_2 \det \left( \boldsymbol{I} + \sigma^{-2} \sum_{k \in \mathcal{S}} \mathbf{H}^{(k)} \boldsymbol{Q}^{(k)}(U) \left(\mathbf{H}^{(k)}\right)^* \right) \right], \forall \mathcal{S} \subseteq \{1, \ldots, K\} \right\}, \qquad (4)$$

*where $\boldsymbol{R} = \left[ R^{(1)} \cdots R^{(K)} \right]^T$ is the rate vector. Additionally, for individual power constraints $\mathrm{Tr}\left(\boldsymbol{Q}^{(k)}(U)\right) \leq P_k$ for $k = 1, \ldots, K$, the capacity region is given by*

$$\mathcal{C}_{MAC}(P_1, \ldots, P_K) = Cl \left( \bigcup_{\{\boldsymbol{Q}^{(k)}(U)\}_{k=1}^K, \mathrm{Tr}(\boldsymbol{Q}^{(k)}(U)) \leq P_k} \right.$$
$$\left. \left\{ \boldsymbol{R} : \sum_{k \in \mathcal{S}} R^{(k)} \leq E_V \left[ \log_2 \det \left( \boldsymbol{I} + \sigma^{-2} \sum_{k \in \mathcal{S}} \mathbf{H}^{(k)} \boldsymbol{Q}^{(k)}(U) \left(\mathbf{H}^{(k)}\right)^* \right) \right], \forall \mathcal{S} \subseteq \{1, \ldots, K\} \right\} \right), \qquad (5)$$

*where $Cl$ denotes the convex closure operation.*

We prove this theorem in the Appendix A.

Fig. 4 shows an example for the capacity regions of two user systems corresponding to $\mathcal{U}_2 = \{u^{(1)}, u^{(2)}\}$ and $\mathcal{U}_4 = \{u^{(1)}, u^{(2)}, u^{(3)}, u^{(4)}\}$ for fixed power allocation $P_1$ and $P_2$. The capacity region is a union of capacity pentagons for fixed covariance matrices that are given by (4).

**Remark** [17–19] With a sum power constraint $\sum_{k=1}^{K} P_k \leq P$, the capacity region is given by

$$\mathcal{C}_{union}(P) = Cl \left( \bigcup_{\sum_{k=1}^{K} P_k \leq P} \mathcal{C}_{MAC}(P_1, \ldots, P_K) \right). \tag{6}$$

using the duality between the MIMO MAC and MIMO BC (broadcasting channel).

## 4 Codebook Design Procedure Under a Sum Power Constraint

In this section, we give a method to design a codebook in a finite rate feedback environment to maximize the sum-rate. The sum-rate can be represented as

$$\begin{aligned}
\sum_{k=1}^{K} R^{(k)} &\leq E_V \left[ \log_2 \det \left( \boldsymbol{I} + \sigma^{-2} \sum_{k=1}^{K} \mathbf{H}^{(k)} \boldsymbol{Q}^{(k)}(U) \left( \mathbf{H}^{(k)} \right)^* \right) \right] \\
&\leq E_V \left[ \max_{\boldsymbol{Q}^{(1)}(U), \ldots, \boldsymbol{Q}^{(K)}(U)} \log_2 \det \left( \boldsymbol{I} + \sigma^{-2} \sum_{k=1}^{K} \mathbf{H}^{(k)} \boldsymbol{Q}^{(k)}(U) \left( \mathbf{H}^{(k)} \right)^* \right) \right] \\
&= \sum_{q=1}^{2^B} \max_{\mathcal{H}_q, \mathrm{Tr}(\boldsymbol{Q}_q) \leq P} E \left[ \log_2 \det \left( \boldsymbol{I} + \sigma^{-2} \sum_{k=1}^{K} \mathbf{H}^{(k)} \boldsymbol{Q}_q^{(k)} \left( \mathbf{H}^{(k)} \right)^* \right) \bigg| \mathbf{H} \in \mathcal{H}_q \right] \mathcal{P}_q, \tag{7}
\end{aligned}$$

where $\mathbf{H} = \begin{bmatrix} \mathbf{H}^{(1)} & \cdots & \mathbf{H}^{(K)} \end{bmatrix}$, $\mathcal{P}_q = P(\mathbf{H} \in \mathcal{H}_q)$, and $\mathcal{H} = \{\mathcal{H}_1, \ldots, \mathcal{H}_{2^B}\}$ is the partition on the channel matrix space $\mathcal{H}$. We let $\boldsymbol{Q}_q^{(k)} = \boldsymbol{Q}^{(k)}(u^{(q)})$ and $\boldsymbol{Q}_q = diag\left(\boldsymbol{Q}_q^{(1)}, \ldots, \boldsymbol{Q}_q^{(K)}\right)$ to simplify the notation, where $u^{(q)}$ is a realization of $U$ and $diag(\cdot, \ldots, \cdot)$ is a block diagonal operator. In this setup, the basestation can simply feed back the covariance index $q \in \{1, 2, \ldots, 2^B\}$ to notify which covariance matrix $\boldsymbol{Q}_q$ should be used to maximize the achievable sum-rate. This means that we can convey this CSIT using a finite rate feedback channel that sends the same $B$ bits of feedback to all users.

We apply Lloyd's algorithm [20] to solve the maximization problem.[1] Because the detailed procedure of Lloyd's algorithm is similar to [20], we only emphasize the difference. In the process of designing the codebook using Lloyd's algorithm, one of the steps is to solve the optimization problem

$$\begin{aligned}
\widetilde{\boldsymbol{Q}}_q &= \operatorname*{argmax}_{\boldsymbol{Q}_q : \mathrm{Tr}(\boldsymbol{Q}_q) \leq P} E \left[ \log_2 \det \left( \boldsymbol{I} + \sigma^{-2} \sum_{k=1}^{K} \mathbf{H}^{(k)} \boldsymbol{Q}_q^{(k)} \left( \mathbf{H}^{(k)} \right)^* \right) \bigg| \mathbf{H} \in \mathcal{H}_q \right] \\
&\stackrel{\triangle}{=} \operatorname*{argmax}_{\boldsymbol{Q}_q : \mathrm{Tr}(\boldsymbol{Q}_q) \leq P} f(\boldsymbol{Q}_q) \tag{8}
\end{aligned}$$

---

[1] As described in [20], Lloyd's algorithm converges to local optimum. As in [20], to obtain a point close to the globally optimal point, we run the same algorithm several times with different initial test channels and initial codebooks. Among the local optimum points, we choose the best point that gives the best capacity value.

for a fixed partition $\mathcal{H} = \{\mathcal{H}_1, \ldots, \mathcal{H}_{2^B}\}$. We illustrate the approximate procedure of solving this optimization problem. By the *heuristic approximation* of [20],

$$f(\boldsymbol{Q}_q) = E\left[\log_2 \det\left(\boldsymbol{I} + \sigma^{-2}\mathbf{H}\boldsymbol{Q}_q\mathbf{H}^*\right) \big| \mathbf{H} \in \mathcal{H}_q\right] \mathcal{P}_q \tag{9}$$

$$\approx \log_2 \det\left(\boldsymbol{I} + \sigma^{-2}E\left[\mathbf{H}^*\mathbf{H}\big|\mathbf{H} \in \mathcal{H}_q\right]\boldsymbol{Q}_q\right) \mathcal{P}_q \tag{10}$$

$$= \log_2 \det\left(\boldsymbol{I} + \sigma^{-2}\boldsymbol{R}_q\boldsymbol{Q}_q\right) \mathcal{P}_q \tag{11}$$

$$= \log_2 \det\left(\boldsymbol{I} + \sigma^{-2}\boldsymbol{S}_q\boldsymbol{Q}_q\boldsymbol{S}_q^*\right) \mathcal{P}_q. \tag{12}$$

Here $\boldsymbol{R}_q = E\left[\mathbf{H}^*\mathbf{H}|\mathbf{H} \in \mathcal{H}_q\right] = \boldsymbol{S}_q^*\boldsymbol{S}_q$ with $\boldsymbol{S}_q = \begin{bmatrix}\boldsymbol{S}_q^{(1)} & \cdots & \boldsymbol{S}_q^{(K)}\end{bmatrix}$, where $\boldsymbol{S}_q^{(k)} \in \mathbb{C}^{M_t K \times M_t}$. Thus, the optimization problem becomes

$$\widetilde{\boldsymbol{Q}}_q = \underset{\boldsymbol{Q}_q:\mathrm{Tr}(\boldsymbol{Q}_q) \leq P}{\mathrm{argmax}} \log_2 \det\left(\boldsymbol{I} + \sigma^{-2}\sum_{k=1}^{K}\boldsymbol{S}_q^{(k)}\boldsymbol{Q}_q^{(k)}\left(\boldsymbol{S}_q^{(k)}\right)^*\right)\mathcal{P}_q. \tag{13}$$

Therefore, the sum power iterative waterfilling algorithm [2] can be applied to this kind of optimization problem by treating the locally averaged $\boldsymbol{S}_q^{(1)}, \ldots, \boldsymbol{S}_q^{(K)}$ as the channel matrices. Solving for $q = 1, \ldots, 2^B$, we can obtain the codebook $\mathcal{Q} = \left\{\widetilde{\boldsymbol{Q}}_1, \ldots, \widetilde{\boldsymbol{Q}}_{2^B}\right\}$. Note that (13) is an approximation of the optimal solution of the original problem.

**Remark** In a typical MAC system, a suboptimal solution can be found using individual power constraints. Indeed, the sum-rate maximization problem under individual power constraints is a special case of the sum-rate maximization problem under the sum power constraint. We can also apply Lloyd's algorithm to design a codebook under individual power constraints. An intermediate step in applying Lloyd's algorithm is to solve the optimization problem

$$\left(\widetilde{\boldsymbol{Q}}_q^{(1)}, \ldots, \widetilde{\boldsymbol{Q}}_q^{(K)}\right) = \underset{\left\{\boldsymbol{Q}_q^{(k)}\right\}_{k=1}^{K}:\mathrm{Tr}\left(\boldsymbol{Q}_q^{(k)}\right)\leq P_k}{\mathrm{argmax}} E\left[\log_2 \det\left(\boldsymbol{I} + \sigma^{-2}\sum_{k=1}^{K}\mathbf{H}^{(k)}\boldsymbol{Q}_q^{(k)}\left(\mathbf{H}^{(k)}\right)^*\right)\bigg|\mathbf{H} \in \mathcal{H}_q\right]. \tag{14}$$

The same procedure as the sum power constraint case gives

$$\left(\widetilde{\boldsymbol{Q}}_q^{(1)}, \ldots, \widetilde{\boldsymbol{Q}}_q^{(K)}\right) = \underset{\left\{\boldsymbol{Q}_q^{(k)}\right\}_{k=1}^{K}:\mathrm{Tr}\left(\boldsymbol{Q}_q^{(k)}\right)\leq P_k}{\mathrm{argmax}} \log_2 \det\left(\boldsymbol{I} + \sigma^{-2}\sum_{k=1}^{K}\boldsymbol{S}_q^{(k)}\boldsymbol{Q}_q^{(k)}\left(\boldsymbol{S}_q^{(k)}\right)^*\right)\mathcal{P}_q, \tag{15}$$

and this problem can be solved by the individual power iterative waterfilling algorithm [1] with the effective channels $\boldsymbol{S}_q^{(1)}, \ldots, \boldsymbol{S}_q^{(K)}$.

## 5 Limited Feedback MAC Beamforming

In this section, we consider beamforming transmission for each user. We obtain the capacity region and derive codebook design methods for the limited feedback environment.

## 5.1 Capacity Analysis

In the beamforming transmission, the received signal is given by

$$\boldsymbol{y}_n = \sum_{k=1}^{K} \boldsymbol{H}_n^{(k)} \boldsymbol{w}_n^{(k)} x_n^{(k)} + \boldsymbol{z}_n = \begin{bmatrix} \boldsymbol{H}_n^{(1)} & \boldsymbol{H}_n^{(2)} & \cdots & \boldsymbol{H}_n^{(K)} \end{bmatrix} \begin{bmatrix} \boldsymbol{w}_n^{(1)} x_n^{(1)} \\ \boldsymbol{w}_n^{(2)} x_n^{(2)} \\ \vdots \\ \boldsymbol{w}_n^{(K)} x_n^{(K)} \end{bmatrix} + \boldsymbol{z}_n, n = 1, \ldots, N, \quad (16)$$

where $\boldsymbol{w}_n^{(k)} \in \mathbb{C}^{M_t}$ is the beamforming vector of the $k$-th users. We assume that $E\left[\left|x_n^{(k)}\right|^2\right] = 1$, which means that the transmission power of the system is given by $\sum_{k=1}^{K} \left\|\boldsymbol{w}_n^{(k)}\right\|^2$.

In the limited feedback system with $B$ bits of feedback, the codebook is given by $\mathcal{W} = \{\boldsymbol{w}_1, \ldots, \boldsymbol{w}_{2^B}\}$, where $\boldsymbol{w}_q = \left[\left(\boldsymbol{w}_q^{(1)}\right)^T \cdots \left(\boldsymbol{w}_q^{(K)}\right)^T\right]^T$. Based on the current feedback $u^{(q)}$, the beamforming vector is selected by $\boldsymbol{w}(u^{(q)}) = \boldsymbol{w}_q$ for some $1 \leq q \leq 2^B$. The capacity region in the limited feedback system is given in the following corollary.

**Corollary 1** *For sum power constraint $\sum_{k=1}^{K} P_k \leq P$ with $\left\|\boldsymbol{w}^{(k)}(u^{(q)})\right\|^2 \leq P_k$ for $k = 1, \ldots, K$, the capacity region subject to a beamforming constraint is*

$$\mathcal{C}_{union}^{beamforming}(P) = Cl \left( \bigcup_{\sum_{k=1}^{K} P_k \leq P} Cl \left( \bigcup_{\{\boldsymbol{w}^{(k)}(u^{(q)})\}_{k=1,q=1}^{K,2^B}, \|\boldsymbol{w}^{(k)}(u^{(q)})\|^2 \leq P_k} \right. \right.$$
$$\left. \left. \left\{ \boldsymbol{R} : \sum_{k \in \mathcal{S}} R^{(k)} \leq E_V \left[ \log_2 \det \left( \boldsymbol{I} + \sigma^{-2} \sum_{k \in \mathcal{S}} \mathbf{H}^{(k)} \boldsymbol{w}^{(k)}(U) \left(\boldsymbol{w}^{(k)}(U)\right)^* \left(\mathbf{H}^{(k)}\right)^* \right) \right], \right. \right. \right.$$
$$\left. \left. \left. \forall \mathcal{S} \subseteq \{1, \ldots, K\} \right\} \right) \right). \quad (17)$$

The proof immediately follows from Theorem 1.

## 5.2 Eigenbeamforming with Lloyd's Algorithm

In this section, we derive the beamforming codebook design method using eigenbeamforming with Lloyd's algorithm. As in Section 4, the sum-rate can be represented as

$$\sum_{k=1}^{K} R^{(k)} \leq \sum_{q=1}^{2^B} \max_{\mathcal{H}_q, \|\boldsymbol{w}_q\|^2 \leq P} E\left[ \log_2 \det \left( \boldsymbol{I} + \sigma^{-2} \sum_{k=1}^{K} \mathbf{H}^{(k)} \boldsymbol{w}_q^{(k)} \left(\boldsymbol{w}_q^{(k)}\right)^* \left(\mathbf{H}^{(k)}\right)^* \right) \bigg| \mathbf{H} \in \mathcal{H}_q \right] \mathcal{P}_q. \quad (18)$$

We also let $\boldsymbol{w}_q^{(k)} = \boldsymbol{w}^{(k)}(u^{(q)})$ for the notational simplicity.

We can also apply Lloyd's algorithm to solve this kind of problem [20]. Similar to Section 4, in the process of solving for the codebook using Lloyd's algorithm, one of the steps is to solve

$$\widetilde{\boldsymbol{w}}_q = \operatorname*{argmax}_{\boldsymbol{w}_q : \|\boldsymbol{w}_q\|^2 \leq P} E\left[ \log_2 \det \left( \boldsymbol{I} + \sigma^{-2} \sum_{k=1}^{K} \mathbf{H}^{(k)} \boldsymbol{w}_q^{(k)} \left(\boldsymbol{w}_q^{(k)}\right)^* \left(\mathbf{H}^{(k)}\right)^* \right) \bigg| \mathbf{H} \in \mathcal{H}_q \right] \mathcal{P}_q$$
$$\stackrel{\triangle}{=} \operatorname*{argmax}_{\boldsymbol{w}_q : \|\boldsymbol{w}_q\|^2 \leq P} f(\boldsymbol{w}_q). \quad (19)$$

This is not a convex optimization problem and is therefore difficult to solve. Thus we approximate this problem by using a lower bound and employing the *heuristic approximation* from [20].

First, we claim that

$$\log_2 \det \left( \boldsymbol{I} + \sigma^{-2} \sum_{k=1}^{K} \mathbf{H}^{(k)} \boldsymbol{w}_q^{(k)} \left(\boldsymbol{w}_q^{(k)}\right)^* \left(\mathbf{H}^{(k)}\right)^* \right) \geq \log_2 \left( 1 + \frac{\sigma^{-2}}{K} \sum_{k=1}^{K} \left(\boldsymbol{w}_q^{(k)}\right)^* \left(\mathbf{H}^{(k)}\right)^* \mathbf{H}^{(k)} \boldsymbol{w}_q^{(k)} \right). \quad (20)$$

This follows from the fact that

$$\begin{aligned}\log_2 \det \left( \boldsymbol{I} + \sigma^{-2} \sum_{k=1}^{K} \mathbf{H}^{(k)} \boldsymbol{w}_q^{(k)} \left(\boldsymbol{w}_q^{(k)}\right)^* \left(\mathbf{H}^{(k)}\right)^* \right) &\geq \log_2 \det \left( \boldsymbol{I} + \sigma^{-2} \mathbf{H}^{(l)} \boldsymbol{w}_q^{(l)} \left(\boldsymbol{w}_q^{(l)}\right)^* \left(\mathbf{H}^{(l)}\right)^* \right) \\ &= \log_2 \left( 1 + \sigma^{-2} \left(\boldsymbol{w}_q^{(l)}\right)^* \left(\mathbf{H}^{(l)}\right)^* \mathbf{H}^{(l)} \boldsymbol{w}_q^{(l)} \right)\end{aligned}$$

for $1 \leq l \leq K$ because $\log_2 \det(\cdot)$ is monotonically increasing in the semi-definite cone. Because $\log_2$ is monotonically increasing function, it is satisfied that

$$\det \left( \boldsymbol{I} + \sigma^{-2} \sum_{k=1}^{K} \mathbf{H}^{(k)} \boldsymbol{w}_q^{(k)} \left(\boldsymbol{w}_q^{(k)}\right)^* \left(\mathbf{H}^{(k)}\right)^* \right) \geq 1 + \sigma^{-2} \left(\boldsymbol{w}_q^{(l)}\right)^* \left(\mathbf{H}^{(l)}\right)^* \mathbf{H}^{(l)} \boldsymbol{w}_q^{(l)}. \quad (21)$$

Adding (21) to both sides for $1 \leq l \leq K$ and dividing by $K$ results in

$$\det \left( \boldsymbol{I} + \sigma^{-2} \sum_{k=1}^{K} \mathbf{H}^{(k)} \boldsymbol{w}_q^{(k)} \left(\boldsymbol{w}_q^{(k)}\right)^* \left(\mathbf{H}^{(k)}\right)^* \right) \geq 1 + \frac{\sigma^{-2}}{K} \sum_{k=1}^{K} \left(\boldsymbol{w}_q^{(k)}\right)^* \left(\mathbf{H}^{(k)}\right)^* \mathbf{H}^{(k)} \boldsymbol{w}_q^{(k)}. \quad (22)$$

Then by the *heuristic approximation* of [20],

$$\begin{aligned}f(\boldsymbol{w}_q) &\geq E\left[\log_2 \left(1 + \frac{\sigma^{-2}}{K} \sum_{k=1}^{K} \left(\boldsymbol{w}_q^{(k)}\right)^* \left(\mathbf{H}^{(k)}\right)^* \mathbf{H}^{(k)} \boldsymbol{w}_q^{(k)}\right) \Big| \mathbf{H} \in \mathcal{H}_q \right] \mathcal{P}_q & (23)\\ &\approx \log_2 \left(1 + \frac{\sigma^{-2}}{K} \sum_{k=1}^{K} \left(\boldsymbol{w}_q^{(k)}\right)^* E\left[\left(\mathbf{H}^{(k)}\right)^* \mathbf{H}^{(k)} \big| \mathbf{H} \in \mathcal{H}_q\right] \boldsymbol{w}_q^{(k)}\right) \mathcal{P}_q & (24)\\ &= \log_2 \left(1 + \frac{\sigma^{-2}}{K} \sum_{k=1}^{K} \left(\boldsymbol{w}_q^{(k)}\right)^* \boldsymbol{R}_q^{(k)} \boldsymbol{w}_q^{(k)}\right) \mathcal{P}_q & (25)\\ &\equiv g(\boldsymbol{w}_q),\end{aligned}$$

where $\boldsymbol{R}_q^{(k)} = E\left[\left(\mathbf{H}^{(k)}\right)^* \mathbf{H}^{(k)} \big| \mathbf{H} \in \mathcal{H}_q\right]$.

Instead of maximizing $f(\boldsymbol{w}_q)$, we utilize the lower bound $g(\boldsymbol{w}_q)$ because it is easier to obtain an analytical solution. Let $\boldsymbol{w}_q^{(k)} = \alpha_q^{(k)} \boldsymbol{v}_q^{(k)}$, $\boldsymbol{w}_q^{(k)} \in \mathbb{C}^{M_t}$, where $\alpha_q^{(k)} = \left\|\boldsymbol{w}_q^{(k)}\right\|$ and $\left\|\boldsymbol{v}_q^{(k)}\right\| = 1$. Then

$$\begin{aligned}g(\boldsymbol{w}_q) &= \log_2 \left(1 + \frac{\sigma^{-2}}{K} \sum_{k=1}^{K} \left(\boldsymbol{w}_q^{(k)}\right)^* \boldsymbol{R}_q^{(k)} \boldsymbol{w}_q^{(k)}\right) \\ &= \log_2 \left(1 + \frac{\sigma^{-2}}{K} \sum_{k=1}^{K} \left(\alpha_q^{(k)}\right)^2 \left(\boldsymbol{v}_q^{(k)}\right)^* \boldsymbol{R}_q^{(k)} \boldsymbol{v}_q^{(k)}\right). & (26)\end{aligned}$$

We can select $\boldsymbol{v}_q^{(k)}$ as the unit-norm eigenvector associated with the maximum eigenvalue of $\boldsymbol{R}_q^{(k)}$. Then,

$$g(\boldsymbol{w}_q) = \log_2 \left(1 + \frac{\sigma^{-2}}{K} \sum_{k=1}^{K} \left(\alpha_q^{(k)}\right)^2 \left(\lambda_q^{(k)}\right)^2 \right), \quad (27)$$

where $\left(\lambda_q^{(k)}\right)^2$ is the maximum eigenvalue of $\boldsymbol{R}_q^{(k)}$. By the Schwartz inequality,

$$g(\boldsymbol{w}_q) \leq \log_2 \left(1 + \frac{\sigma^{-2}}{K}\sqrt{\sum_{k=1}^{K}\left(\left(\alpha_q^{(k)}\right)^2\right)^2}\sqrt{\sum_{k=1}^{K}\left(\left(\lambda_q^{(k)}\right)^2\right)^2}\right). \tag{28}$$

Equality holds if and only if

$$\left(\alpha_q^{(k)}\right)^2 = a\left(\lambda_q^{(k)}\right)^2 \tag{29}$$

for some positive constant $a$. Since $\sum_{k=1}^{K}\left(\alpha_q^{(k)}\right)^2 = P$,

$$a = \frac{P}{\sum_{k=1}^{K}\left(\lambda_q^{(k)}\right)^2}. \tag{30}$$

Therefore

$$\alpha_q^{(k)} = \frac{\sqrt{P}\lambda_q^{(k)}}{\sqrt{\sum_{k=1}^{K}\left(\lambda_q^{(k)}\right)^2}} \tag{31}$$

and

$$\boldsymbol{w}_q^{(k)} = \frac{\sqrt{P}\lambda_q^{(k)}}{\sqrt{\sum_{k=1}^{K}\left(\lambda_q^{(k)}\right)^2}}\boldsymbol{v}_q^{(k)}, \tag{32}$$

where $\left(\lambda_q^{(k)}\right)^2$ is the maximum eigenvalue of $\boldsymbol{R}_q^{(k)}$ and $\boldsymbol{v}_q^{(k)}$ is the corresponding unit-norm eigenvector.

The proposed method can additionally be validated by the papers [4, 5, 9], which state that when the covariance feedback is available at the transmitters the optimum beamforming vectors can be found as the eigenvector associated with the maximum eigenvalue of the covariance channel matrix of each user.

## 5.3 Grassmannian Design

In this section, we develop the Granssmannian codebook by solving the Grassmannian subspace packing problem [21] with random power allocation. References [22–25] dealt with a random beamforming codebook by employing random vector quantization (RVQ) in single user systems. This is based on the idea that the right singular vector of the channel which maximizes the capacity is random and isotropically distributed [24] under the assumption of an independent and identically distributed (i.i.d.) complex Gaussian channel matrix. In the same sense, in the MAC beamforming problem, each user's channel is i.i.d., and therefore we can assume that the power allocation of each user is randomly distributed according to sum power constraint. In this correspondence, the beamforming direction of each user will be generated in the Grassmannian sense [21]. The codebook design procedure will be described in the following.

### 5.3.1 Codebook design

Let $\mathcal{W} = \{\boldsymbol{w}_1, \ldots, \boldsymbol{w}_{2^B}\}$ be the beamforming codebook, where $\boldsymbol{w}_q = \left[\left(\boldsymbol{w}_q^{(1)}\right)^T \cdots \left(\boldsymbol{w}_q^{(K)}\right)^T\right]^T$. Here the sum power constraint is given by $\|\boldsymbol{w}_q\|^2 \leq P$ for $q = 1, \ldots, 2^B$. Then the sum-rate capacity is

$$\begin{aligned} C_{sum}^{beamforming} &= E_{\mathbf{H},\mathcal{W}} \left[\max_{\boldsymbol{w}_q \in \mathcal{W}} \log_2 \det \left(\boldsymbol{I} + \sigma^{-2} \sum_{k=1}^{K} \mathbf{H}^{(k)} \boldsymbol{w}_q^{(k)} \left(\boldsymbol{w}_q^{(k)}\right)^* \left(\mathbf{H}^{(k)}\right)^*\right)\right] \\ &= E_{\mathbf{H},\mathcal{W}} \left[\max_{\boldsymbol{w}_q \in \mathcal{W}} \log_2 \det \left(\boldsymbol{I} + \sigma^{-2} \mathbf{H} \boldsymbol{W}_q \boldsymbol{W}_q^* \mathbf{H}^*\right)\right], \end{aligned} \qquad (33)$$

where

$$\boldsymbol{W}_q = \begin{bmatrix} \boldsymbol{w}_q^{(1)} & \boldsymbol{0}_{M_t} & \cdots & \boldsymbol{0}_{M_t} \\ \boldsymbol{0}_{M_t} & \boldsymbol{w}_q^{(2)} & \cdots & \boldsymbol{0}_{M_t} \\ \vdots & \vdots & \ddots & \vdots \\ \boldsymbol{0}_{M_t} & \boldsymbol{0}_{M_t} & \cdots & \boldsymbol{w}_q^{(K)} \end{bmatrix} \qquad (34)$$

is a beamforming matrix obtained by augmenting with each user's beamforming vector with the zero column vector of length $M_t$, $\boldsymbol{0}_{M_t}$. We can see that (33) is the capacity formula for the single user MIMO system with the covariance matrix $\boldsymbol{W}_q \boldsymbol{W}_q^*$ of the constraint that $\boldsymbol{W}_q$ is block diagonal when $\{\boldsymbol{W}_1, \ldots, \boldsymbol{W}_{2^B}\}$ is a codebook. Note that $\boldsymbol{W}_q \in \mathbb{C}^{M_t K \times K}$ has orthogonal columns, but it is not unitary because $\boldsymbol{W}_q^* \boldsymbol{W}_q \neq \alpha \boldsymbol{I}$, where $\alpha$ is some non-negative constant chosen according to sum power constraint. In the unitary case of single user MIMO systems, references [22, 24] designed random codebooks by RVQ and reference [21] designed Grassmannian codebooks by Grassmannian subspace packing with the *Fubini-Study distance*. However, these codebook design methods are not directly applicable to our problem because our augmented beamforming matrix is not unitary.

In this section, we will assume that $\boldsymbol{w}_q^{(k)} = \alpha_q^{(k)} \boldsymbol{v}_q^{(k)}$, where $\alpha_q^{(k)}$ is the power allocation factor for the user $k$ and $\boldsymbol{v}_q^{(k)}$ is the normalized unit-norm beamforming vector for the user $k$. Here we assume that $\sum_{k=1}^{K} \left|\alpha_q^{(k)}\right|^2 = P$ for sum power constraint $P$. The augmented beamforming matrix is given by

$$\boldsymbol{W}_q = \boldsymbol{B}_q \boldsymbol{V}_q, \qquad (35)$$

where $\boldsymbol{B}_q = \boldsymbol{A}_q \otimes \boldsymbol{I}$ with $\boldsymbol{A}_q = diag\left(\alpha_q^{(1)}, \ldots, \alpha_q^{(K)}\right)$ and

$$\boldsymbol{V}_q = \begin{bmatrix} \boldsymbol{v}_q^{(1)} & \boldsymbol{0}_{M_t} & \cdots & \boldsymbol{0}_{M_t} \\ \boldsymbol{0}_{M_t} & \boldsymbol{v}_q^{(2)} & \cdots & \boldsymbol{0}_{M_t} \\ \vdots & \vdots & \ddots & \vdots \\ \boldsymbol{0}_{M_t} & \boldsymbol{0}_{M_t} & \cdots & \boldsymbol{v}_q^{(K)} \end{bmatrix}. \qquad (36)$$

To maximize the sum-rate, it suffices to maximize the average determinant

$$
\begin{aligned}
C_{sum}^{beamforming} &\doteq E_{\mathbf{H},\mathcal{W}}\left[\max_{\boldsymbol{w}_q\in\mathcal{W}} \det\left(\boldsymbol{I}+\sigma^{-2}\sum_{k=1}^{K}\mathbf{H}^{(k)}\boldsymbol{w}_q^{(k)}\left(\boldsymbol{w}_q^{(k)}\right)^*\left(\mathbf{H}^{(k)}\right)^*\right)\right] \\
&= E_{\mathbf{H},\mathcal{W}}\left[\max_{\boldsymbol{w}_q\in\mathcal{W}} \det\left(\boldsymbol{I}+\sigma^{-2}\mathbf{H}\boldsymbol{B}_q\boldsymbol{V}_q\boldsymbol{V}_q^*\boldsymbol{B}_q^*\mathbf{H}^*\right)\right] \\
&= E_{\mathbf{H},\mathcal{W}}\left[\max_{\boldsymbol{w}_q\in\mathcal{W}} \det\left(\boldsymbol{I}+\sigma^{-2}\mathbf{H}_A\boldsymbol{V}_q\boldsymbol{V}_q^*\mathbf{H}_A^*\right)\right],
\end{aligned}
\quad (37)
$$

where $\doteq$ means the equivalence of the optimization and $\mathbf{H}_A = \mathbf{H}\boldsymbol{B}_q$. Thus, the codebook design problem can be viewed as the unitary codebook design of $\mathcal{V} \triangleq \{\boldsymbol{V}_1,\ldots,\boldsymbol{V}_{2^B}\}$ with the equivalent channel $\mathbf{H}_A$ [21]. This problem can be solved by Grassmannian subspace packing using the *Fubini-Study distance*. The *Fubini-Study distance* is given by [21]

$$
\begin{aligned}
d_{FS}(\boldsymbol{V}_q,\boldsymbol{V}_r) &= \arccos|\det(\boldsymbol{V}_q^*\boldsymbol{V}_r)| \\
&= \arccos\left(\prod_{k=1}^{K}\left|\left(\boldsymbol{v}_q^{(k)}\right)^*\boldsymbol{v}_r^{(k)}\right|\right).
\end{aligned}
\quad (38)
$$

Let $\delta_{FS}(\mathcal{V}) = \min_{1\leq q<r\leq 2^B} d_{FS}(\boldsymbol{V}_q,\boldsymbol{V}_r)$. By [21], the codebook $\mathcal{V}$ can be constructed by maximizing the minimum distance $\delta_{FS}(\mathcal{V})$. A detailed procedure to maximize $\delta_{FS}(\mathcal{V})$ is given in the Appendix B using Lloyd's algorithm. $\boldsymbol{B}_q$ can be randomly generated according to the sum power constraint, and the final beamforming vector can be obtained by $\boldsymbol{w}_q^{(k)} = \alpha_q^{(k)}\boldsymbol{v}_q^{(k)}$, $k=1,\ldots,K$, $q=1,\ldots,2^B$. The codebook can be constructed as $\mathcal{W} = \{\boldsymbol{w}_1,\cdots,\boldsymbol{w}_{2^B}\}$, where $\boldsymbol{w}_q = \left[\left(\boldsymbol{w}_q^{(1)}\right)^T \cdots \left(\boldsymbol{w}_q^{(K)}\right)^T\right]^T$.

### 5.3.2 Codebook Design in Correlated Fading

In this section, we discuss how the codebook design can be modified for a correlated fading channel. We assume that the receiver experiences an uncorrelated fading environment [4,5]. Under a Kronecker correlated channel model, the general channel model of the user $k$ is given by [4,5,26–28]

$$
\mathbf{H}^{(k)} = \mathbf{H}_w^{(k)}\left(\boldsymbol{R}_t^{(k)}\right)^{1/2}, \quad (39)
$$

where $\boldsymbol{R}_t^{(k)}$ is covariance matrix of transmit antennas for user $k$. The $M_r \times M_t$ random matrix $\mathbf{H}_w^{(k)}$ is i.i.d. circular symmetric complex Gaussian with zero-mean and unit-variance.

In this channel correlation environment, the statistical beamforming is introduced to achieve the sum-rate capacity [4,5]. The statistical beamforming matrix under the assumption that equal power is allocated for each user is given by [4,5]

$$
\boldsymbol{V}_{stat} = \begin{bmatrix} \boldsymbol{v}_{stat}^{(1)} & \boldsymbol{0}_{M_t} & \cdots & \boldsymbol{0}_{M_t} \\ \boldsymbol{0}_{M_t} & \boldsymbol{v}_{stat}^{(2)} & \cdots & \boldsymbol{0}_{M_t} \\ \vdots & \vdots & \ddots & \vdots \\ \boldsymbol{0}_{M_t} & \boldsymbol{0}_{M_t} & \cdots & \boldsymbol{v}_{stat}^{(K)} \end{bmatrix}, \quad (40)
$$

where $\boldsymbol{v}_{stat}^{(k)}$ is the eigenvector associated with the maximum eigenvalue of $\boldsymbol{R}_t^{(k)}$. The statistical beamforming approach considered in [4, 5] has excellent performance in highly correlated channel, but suffers when correlation is moderate. We can improve the sum-rate performance of statistical beamforming in a limited feedback scenario by leveraging a modified codebook design [29]. This follows from the fact that the minimum distance of the codebook $\mathcal{V} = \{\boldsymbol{V}_1, \ldots, \boldsymbol{V}_{2^B}\}$ from Lloyd's algorithm is the same as the the minimum distance of the codebook $\widetilde{\mathcal{V}} = \{\boldsymbol{U}\boldsymbol{V}_1, \ldots, \boldsymbol{U}\boldsymbol{V}_{2^B}\}$, where $\boldsymbol{U}$ is an arbitrary unitary matrix. We can design our codebook such that $\boldsymbol{V}_{stat} = \boldsymbol{U}\boldsymbol{V}_1$ and the corresponding codebook is $\widetilde{\mathcal{V}} = \{\boldsymbol{U}\boldsymbol{V}_1, \ldots, \boldsymbol{U}\boldsymbol{V}_{2^B}\}$.

In highly correlated channel environments, the proposed rotated codebook in the limited feedback environment has similar performance to the statistical beamforming. However, in moderate correlated channel environment, the proposed codebook has the better sum-rate performance than the statistical beamforming. We will see this increase of sum-rate in the numerical simulation section.

# 6 Numerical Simulation

In this section, we present numerical simulation results that show the sum-rate capacity performance of the proposed codebooks using Monte Carlo simulations. We call the multiple access system that has $K$ users, $M_t$ transmit antennas, and $M_r$ receive antennas a $(K, M_t, M_r)$ system.

**Covariance Codebook**

In this correspondence, we will compare the performance of the proposed system against a system with full CSI feedback and a system with no feedback. By no feedback, we mean that we allocate some scaled identity matrix to each data covariance matrix according to the power constraint. By full CSI feedback, we mean that we use the well-known iterative waterfilling [2] under a sum power constraint when the channel state information is known perfectly at all transmitters.

Fig. 5 shows the simulation result for a $(2, 2, 4)$ system under sum power constraint. The upper bound is the sum-rate performance when the users know the full CSI. The lower bound is the sum-rate performance when the users have no knowledge of the CSI. As we can see in this figure, as the number of feedback bits increases, the sum-rate capacity gain is also increased.

**Beamforming Codebook**

We show the performance of the eigenbeamforming with Lloyd's algorithm and the Grassmannian codebook with random power allocation. For the Grassmannian packing, to compare the performance, we simulated the random codebook with random power allocation.

First, we show the performance of eigenbeamforming with Lloyd's algorithm under sum power constraint. We simulated the system assuming each user receives 1 to 4 feedback bits to identify its beamforming vector. Fig. 6 shows the performance of the proposed eigenbeamforming of a (5,3,3) system. As the number of feedback bits increases, the sum-rate capacity also increases. Using 4 bits instead of 1 bit provides around a 1 dB gain. Compared to no feedback system, using 4 bits provides around a 2 dB SNR gain.

Fig. 7 shows the sum-rate performance of the Grassmannian codebook and the random codebook for a (2,2,3) system. From the figure, we can see that the two beamforming methods work well. Using 3 bits instead of 1 bit provides around a 1.5 dB SNR gain. We can also see that the sum-rate performance of the Grassmannian codebook is better than that of the random codebook. This is because the minimum distance of the Grassmannian codebook is larger than that of the random codebook. We can expect that the larger the minimum distance, the higher the obtained performance gain.

Fig. 8 shows the sum-rate performance of our Grasmannian codebook and statistical beamforming in correlated channel for a (2,2,3) system. We used the correlated channel model such that the eigenvalues of $\boldsymbol{R}_t$ are made be 1.2 and 0.8 for each user. We can see that using 3 bit feedback information instead of statistical beamforming has an SNR advantage of around 1.5 dB. Note that the performance gain would be reduced if the transmitter spatial correlation increased.

Fig. 9 shows the relative performance comparison between the covariance codebook, the eigenbeamforming codebook and the Grassmannian beamforming codebook for a (2,2,4) system with 2 bit feedback. The performance of the full CSI is included as a reference. As expected, the covariance codebook has the best performance of all the codebook schemes and especially at high SNR the performance is close to the optimal iterative waterfilling solution which has the same slope of increasing according to SNR. Also for the beamforming schemes, as expected we do not have a multiplexing gain compared to the covariance codebook.

Fig. 10 shows the performance between the covariance codebook and time-division multiple access (TDMA) for a (2,2,4) system with 2 bit feedback. We can see that the proposed covariance codebook has similar performance compared to the TDMA scheme at low SNR. However, at high SNR, the performance gap becomes large. The second figure shows the ratio defined in [30, eq. (8)] between the covariance codebook and TDMA scheme. We can also verify the performance increases at high SNR. We can also observe the same performance trend as [30, Fig. 3]

# 7  Conclusion

In this correspondence, we proved a coding theorem for the MIMO MAC with finite cardinality feedback and proposed a codebook (a set of covariance matrices) design algorithm with the spatial waterfilling design using Lloyd's algorithm. Compared to a no feedback system, a performance enhancement is observed even for a small number of feedback bits. Compared to full feedback, the proposed covariance codebook achieves comparable performance. We also derived the capacity region and two codebook design methods based on eigenbeamforming and Grassmannian subspace packing when we use beamforming for each user. Simulation results show that we can achieve more sum-rate capacity gains compared to the random beamforming codebook. Also for correlated channels, the proposed Grassmannian beamforming system has better sum-rate performance than the statistical beamforming codebook.

## Appendix A: Proof of Theorem 1

We will prove this theorem in several steps.

**Lemma 1** Let $\mathbf{x}_n(\mathcal{S}) = \left[\left(\mathbf{x}_n^{(k_1)}\right)^T \cdots \left(\mathbf{x}_n^{(k_{|\mathcal{S}|})}\right)^T\right]^T$ with $\mathcal{S} = \{k_1, \ldots, k_{|\mathcal{S}|}\} \subseteq \{1, \ldots, K\}$. $\mathbf{x}_n(\mathcal{S}^C)$ can be defined in the same way with $\mathcal{S}^C$. Then

$$I\left(\mathbf{x}_n\left(\mathcal{S}\right); \mathbf{y}_n | V_n, \mathbf{x}_n\left(\mathcal{S}^C\right)\right) \leq E_{V_n}\left[\log_2 \det\left(\mathbf{I} + \sigma^{-2} \sum_{k \in \mathcal{S}} \mathbf{H}_n^{(k)} \mathbf{Q}^{(k)}(U_n)\left(\mathbf{H}_n^{(k)}\right)^*\right)\right] \equiv C(\mathcal{S}). \tag{41}$$

**Proof**

$$\begin{aligned}
I\left(\mathbf{x}_n\left(\mathcal{S}\right); \mathbf{y}_n | V_n, \mathbf{x}_n\left(\mathcal{S}^C\right)\right) &= h\left(\mathbf{y}_n | V_n, \mathbf{x}_n\left(\mathcal{S}^C\right)\right) - h\left(\mathbf{y}_n | \mathbf{x}_n\left(\mathcal{S}\right), V_n, \mathbf{x}_n\left(\mathcal{S}^C\right)\right) \\
&= h\left(\mathbf{y}_n | V_n, \mathbf{x}_n\left(\mathcal{S}^C\right)\right) - h\left(\mathbf{z}_n\right) \\
&\leq E_{V_n}\left[\log_2 \det\left(\mathbf{I} + \sigma^{-2} E\left[\mathbf{H}_n\left(\mathcal{S}\right) \mathbf{x}_n\left(\mathcal{S}\right) \mathbf{x}_n^*\left(\mathcal{S}\right) \mathbf{H}_n^*\left(\mathcal{S}\right) | V_n\right]\right)\right] \tag{42} \\
&= E_{V_n}\left[\log_2 \det\left(\mathbf{I} + \sigma^{-2} \sum_{k \in \mathcal{S}} \mathbf{H}_n^{(k)} E\left[\mathbf{x}_n^{(k)}\left(\mathbf{x}_n^{(k)}\right)^* | U_n\right] \left(\mathbf{H}_n^{(k)}\right)^*\right)\right] \\
&= E_{V_n}\left[\log_2 \det\left(\mathbf{I} + \sigma^{-2} \sum_{k \in \mathcal{S}} \mathbf{H}_n^{(k)} \mathbf{Q}^{(k)}(U_n)\left(\mathbf{H}_n^{(k)}\right)^*\right)\right], \tag{43}
\end{aligned}$$

where $\mathbf{H}_n\left(\mathcal{S}\right) = \left[\mathbf{H}_n^{(k_1)} \cdots \mathbf{H}_n^{(k_{|\mathcal{S}|})}\right]$ and $E\left[\mathbf{x}_n^{(k)}\left(\mathbf{x}_n^{(k)}\right)^* | U_n\right] = \mathbf{Q}^{(k)}(U_n)$. In (42), we use the maximum entropy property because the covariance matrix of $\mathbf{y}_n$ given $V_n$ and $\mathbf{x}_n(\mathcal{S}^C)$ is $E\left[\mathbf{H}_n\left(\mathcal{S}\right) \mathbf{x}_n\left(\mathcal{S}\right) \mathbf{x}_n^*\left(\mathcal{S}\right) \mathbf{H}_n^*\left(\mathcal{S}\right) | V_n\right] + \sigma^2 \mathbf{I}$ [16]. ∎

**Lemma 2** Let $\mathbf{T} = \left[\left(\mathbf{T}^{(1)}\right)^T \cdots \left(\mathbf{T}^{(K)}\right)^T\right]^T$ and $\mathbf{T}(\mathcal{S}) = \left[\left(\mathbf{T}^{(k_1)}\right)^T \cdots \left(\mathbf{T}^{(k_{|\mathcal{S}|})}\right)^T\right]^T$. $\mathbf{T}(\mathcal{S}^C)$ is defined in the same way. Then

$$I\left(\mathbf{T}\left(\mathcal{S}\right); \mathbf{T}\left(\mathcal{S}^C\right) | \mathbf{y}, V\right) = I\left(\mathbf{x}\left(\mathcal{S}\right); \mathbf{x}\left(\mathcal{S}^C\right) | \mathbf{y}, V\right). \tag{44}$$

**Proof** We know that $\mathbf{T}^{(k)} = \begin{bmatrix} \mathbf{t}_1^{(k)} & \cdots & \mathbf{t}_{|\mathcal{U}|}^{(k)} \end{bmatrix}$, where $\mathbf{t}_i^{(k)} \in \mathcal{X}^{M_t}$. $\mathbf{t}_j(\mathcal{S})$ is defined as $\mathbf{t}_j(\mathcal{S}) = \begin{bmatrix} \left(\mathbf{t}_j^{(k_1)}\right)^T \cdots \left(\mathbf{t}_j^{(k_{|\mathcal{S}|})}\right)^T \end{bmatrix}^T$, *i.e.* the $j$-th column of $\mathbf{T}(\mathcal{S})$. $\mathbf{t}_j(\mathcal{S}^C)$ is defined by the same way. We define a function $f : \mathcal{U} \to \{1, \ldots, |\mathcal{U}|\}$, where $f(U)$ is a bijection. We also define $\mathbf{T}(\mathcal{S}, U) \equiv \mathbf{t}_{f(U)}(\mathcal{S})$ and

$$\mathbf{T}\left(\mathcal{S}, U^C\right) \equiv \begin{bmatrix} \mathbf{t}_1(\mathcal{S}) & \cdots & \mathbf{t}_{f(U)-1}(\mathcal{S}) & \mathbf{t}_{f(U)+1}(\mathcal{S}) & \cdots & \mathbf{t}_{|\mathcal{U}|}(\mathcal{S}) \end{bmatrix}. \tag{45}$$

$\mathbf{T}\left(\mathcal{S}^C, U\right)$ and $\mathbf{T}\left(\mathcal{S}^C, U^C\right)$ can be defined in the same way.

Then

$$I\left(\mathbf{T}(\mathcal{S}); \mathbf{T}(\mathcal{S}^C) | \mathbf{y}, V\right) = I\left(\mathbf{T}(\mathcal{S}, U^C), \mathbf{T}(\mathcal{S}, U); \mathbf{T}(\mathcal{S}^C, U^C), \mathbf{T}(\mathcal{S}^C, U) | \mathbf{y}, V\right) \tag{46}$$

$$= I\left(\mathbf{T}(\mathcal{S}^C, U); \mathbf{T}(\mathcal{S}, U) | \mathbf{y}, V\right)$$

$$+ I\left(\mathbf{T}(\mathcal{S}^C, U); \mathbf{T}(\mathcal{S}, U^C) | \mathbf{T}(\mathcal{S}, U), \mathbf{y}, V\right)$$

$$+ I\left(\mathbf{T}(\mathcal{S}^C, U^C); \mathbf{T}(\mathcal{S}, U) | \mathbf{T}(\mathcal{S}^C, U), \mathbf{y}, V\right)$$

$$+ I\left(\mathbf{T}(\mathcal{S}^C, U^C); \mathbf{T}(\mathcal{S}, U^C) | \mathbf{T}(\mathcal{S}^C, U), \mathbf{T}(\mathcal{S}, U), \mathbf{y}, V\right)$$

$$= I\left(\mathbf{T}(\mathcal{S}^C, U); \mathbf{T}(\mathcal{S}, U) | \mathbf{y}, V\right) \tag{47}$$

$$= I\left(\mathbf{x}(\mathcal{S}); \mathbf{x}(\mathcal{S}^C) | \mathbf{y}, V\right). \tag{48}$$

Eqn. (47) follows from independence of data for the set $\mathcal{S}$ and $\mathcal{S}^C$ according to $U$ and $U^C$. Eqn. (48) follows from $\mathbf{x}(\mathcal{S}) = \mathbf{T}(\mathcal{S}, U)$ and $\mathbf{x}(\mathcal{S}^C) = \mathbf{T}(\mathcal{S}^C, U)$. Here, we used $I(A, B; C, D) = I(D; B) + I(D; A|B) + I(C; B|D) + I(C; A|D, B)$. This can easily be seen by the chain rule. ∎

Now we prove Theorem 1.

**Proof** We first prove (4).

*Achievability*: We follow the procedure of [31] with the equivalent channel in Fig. 3.

Codebook generation: Fix $p\left(\boldsymbol{T}^{(1)}, \ldots, \boldsymbol{T}^{(K)}\right) = \prod_{k=1}^{K} p\left(\boldsymbol{T}^{(k)}\right)$. We generate $2^{NR^{(k)}}$ independent codewords $\left\{\mathbf{T}^{(k)}\left(\omega^{(k)}\right)\right\}_1^N$, $\omega^{(k)} \in \left\{1, 2, \ldots, 2^{NR^{(k)}}\right\}$ with respect to $\prod_{n=1}^{N} p\left(\boldsymbol{T}_n^{(k)}\right)$, where

$$\left\{\mathbf{T}^{(k)}\left(\omega^{(k)}\right)\right\}_1^N = \left\{\mathbf{T}_1^{(k)}\left(\omega^{(k)}\right), \ldots, \mathbf{T}_N^{(k)}\left(\omega^{(k)}\right)\right\}. \tag{49}$$

These codewords form the codebook, which is revealed to the transmitters and the receiver.

Encoding: To send index $\omega^{(k)}$, sender $k$ sends the codeword $\mathbf{T}_1^{(k)}\left(\omega^{(k)}\right), \ldots, \mathbf{T}_N^{(k)}\left(\omega^{(k)}\right)$ through the equivalent channel $p(\boldsymbol{y}_1^N, v_1^N | \boldsymbol{T}_1^N)$.

Decoding: Let $A_\epsilon^{(N)}$ denote the set of typical $\left(\{T^{(1)}\}_1^N, \ldots, \{T^{(K)}\}_1^N, \boldsymbol{y}_1^N, v_1^N\right)$ sequences. The properties of $A_\epsilon^{(N)}$ are given in [31]. The receiver chooses the $K$-tuple $(\omega^{(1)}, \ldots, \omega^{(K)})$ such that

$$\left(\{T^{(1)}(\omega^{(1)})\}_1^N, \ldots, \{T^{(K)}(\omega^{(K)})\}_1^N, \boldsymbol{y}_1^N, v_1^N\right) \in A_\epsilon^{(N)} \qquad (50)$$

if such a $K$-tuple $(\omega^{(1)}, \ldots, \omega^{(K)})$ exists and is unique; otherwise, an error is declared.

Analysis of the probability error: We can assume that $(\omega^{(1)}, \ldots, \omega^{(K)}) = (1, \ldots, 1)$. We have an error if either the correct codewords are not typical with the received sequence or there is a pair of incorrect codewords that are typical with the received sequence. Define the events

$$E_{\omega^{(1)}, \ldots, \omega^{(K)}} = \left\{\left(\{T^{(1)}(\omega^{(1)})\}_1^N, \ldots, \{T^{(K)}(\omega^{(K)})\}_1^N, \boldsymbol{y}_1^N, v_1^N\right) \in A_\epsilon^{(N)}\right\} \qquad (51)$$

and

$$E_0^{(k)} = \left\{\{T^{(k)}(1)\}_1^N : \left|\operatorname{Tr}\left(\frac{1}{N}\sum_{n=1}^N t_{f(U_n)}^{(k)}(1)\left(t_{f(U_n)}^{(k)}(1)\right)^*\right) - P_k\right| > \epsilon_k\right\} \qquad (52)$$

for $U_n \in \mathcal{U}$. Then

$$P_e^{(N)} = P\left(\left(\bigcup_{k=1}^K E_0^{(k)}\right) \bigcup E_{1,\ldots,1}^C \bigcup \left(\bigcup_{(\omega^{(1)},\ldots,\omega^{(K)}) \neq (1,\ldots,1)} E_{\omega^{(1)},\ldots,\omega^{(K)}}\right)\right)$$

$$\leq \sum_{k=1}^K P\left(E_0^{(k)}\right) + P\left(E_{1,\ldots,1}^C\right) + \sum_{\mathcal{S} \subseteq \{1,\ldots,K\}} P\left(E_{\omega(\mathcal{S}) \neq (1,\ldots,1), \omega(\mathcal{S}^C) = (1,\ldots,1)}\right), \qquad (53)$$

where $\omega(\mathcal{S}) = \{\omega^{(k_1)}, \ldots, \omega^{(k_{|\mathcal{S}|})}\}$. We can prove that $P\left(E_{1,\ldots,1}^C\right) \leq \epsilon$ by the property of $\epsilon$-typical sequence. We can also prove that $P\left(E_0^{(k)}\right) \leq \epsilon$ by the law of large numbers. If we define

$$\mathbf{T}_1^N\left(\mathcal{S}, \omega^{(k_1)}, \ldots, \omega^{(k_{|\mathcal{S}|})}\right) = \left\{\{T^{(k_1)}(\omega^{(k_1)})\}_1^N, \ldots, \{T^{(k_{|\mathcal{S}|})}(\omega^{(k_{|\mathcal{S}|})})\}_1^N\right\}, \qquad (54)$$

then

$$P\left(E_{\omega(\mathcal{S})=(\omega^{(k_1)},\ldots,\omega^{(k_{|\mathcal{S}|})}), \omega(\mathcal{S}^C)=(1,\ldots,1)}\right)$$
$$= P\left(\left(\mathbf{T}_1^N\left(\mathcal{S}, \omega^{(k_1)}, \ldots, \omega^{(k_{|\mathcal{S}|})}\right), \mathbf{T}_1^N\left(\mathcal{S}^C, 1, \ldots, 1\right), \mathbf{y}_1^N, V_1^N\right) \in A_\epsilon^{(N)}\right)$$
$$= \sum_{(\{T^{(1)}\}_1^N, \ldots, \{T^{(K)}\}_1^N, \boldsymbol{y}_1^N, v_1^N) \in A_\epsilon^{(N)}} p\left(\mathbf{T}_1^N(\mathcal{S})\right) p\left(\mathbf{T}_1^N(\mathcal{S}^C), \boldsymbol{y}_1^N, v_1^N\right)$$
$$\leq |A_\epsilon^{(N)}| 2^{-N(h(\mathbf{T}(\mathcal{S}))-\epsilon)} 2^{-N(h(\mathbf{T}(\mathcal{S}^C),\mathbf{y},V)-\epsilon)}$$
$$\leq 2^{N(h(\mathbf{T}(\mathcal{S}),\mathbf{T}(\mathcal{S}^C),\mathbf{y},V)+\epsilon)} 2^{-N(h(\mathbf{T}(\mathcal{S}))-\epsilon)} 2^{-N(h(\mathbf{T}(\mathcal{S}^C),\mathbf{y},V)-\epsilon)}$$
$$= 2^{-N(h(\mathbf{T}(\mathcal{S}))+h(\mathbf{T}(\mathcal{S}^C),\mathbf{y},V)-h(\mathbf{T}(\mathcal{S}),\mathbf{T}(\mathcal{S}^C),\mathbf{y},V)-3\epsilon)}$$
$$= 2^{-N(I(\mathbf{T}(\mathcal{S});\mathbf{T}(\mathcal{S}^C),\mathbf{y},V)-3\epsilon)}$$
$$= 2^{-N(I(\mathbf{T}(\mathcal{S});\mathbf{y}|\mathbf{T}(\mathcal{S}^C),V)-3\epsilon)}, \qquad (55)$$

where $\mathbf{T}_1^N(\mathcal{S}) = \{\mathbf{T}_1(\mathcal{S}), \ldots, \mathbf{T}_N(\mathcal{S})\}$. The last equality follows from

$$I\left(\mathbf{T}\left(\mathcal{S}\right); \mathbf{T}\left(\mathcal{S}^C\right), \mathbf{y}, V\right) = I\left(\mathbf{T}\left(\mathcal{S}\right); \mathbf{T}\left(\mathcal{S}^C\right), V\right) + I\left(\mathbf{T}\left(\mathcal{S}\right); \mathbf{y}|\mathbf{T}\left(\mathcal{S}^C\right), V\right) \tag{56}$$

and $I\left(\mathbf{T}\left(\mathcal{S}\right); \mathbf{T}\left(\mathcal{S}^C\right), V\right) = 0$ since $\mathbf{T}\left(\mathcal{S}\right)$ and $\mathbf{T}\left(\mathcal{S}^C\right)$ are independent.

Therefore

$$P_e^{(N)} \leq (K+1)\epsilon + \sum_{\mathcal{S} \subseteq \{1,\ldots,K\}} 2^{N\left(R(\mathcal{S}) - I\left(\mathbf{T}(\mathcal{S}); \mathbf{y}|\mathbf{T}(\mathcal{S}^C), V\right) + 3\epsilon\right)}, \tag{57}$$

where $R\left(\mathcal{S}\right) = \sum_{k \in \mathcal{S}} R^{(k)}$. Therefore, if $R\left(\mathcal{S}\right) < I\left(\mathbf{T}\left(\mathcal{S}\right); \mathbf{y}|\mathbf{T}\left(\mathcal{S}^C\right), V\right)$, $P_e^{(N)} \to 0$ as $N \to \infty$.

To complete the proof, it suffices to show that

$$I\left(\mathbf{T}\left(\mathcal{S}\right); \mathbf{y}|\mathbf{T}\left(\mathcal{S}^C\right), V\right) \leq I\left(\mathbf{x}\left(\mathcal{S}\right); \mathbf{y}|\mathbf{x}\left(\mathcal{S}^C\right), V\right). \tag{58}$$

This is because

$$I\left(\mathbf{T}\left(\mathcal{S}\right); \mathbf{y}|\mathbf{T}\left(\mathcal{S}^C\right), V\right)$$

$$= I\left(\mathbf{T}\left(\mathcal{S}\right); \mathbf{y}|V\right) - I\left(\mathbf{T}\left(\mathcal{S}\right); \mathbf{T}\left(\mathcal{S}^C\right)|V\right) + I\left(\mathbf{T}\left(\mathcal{S}\right); \mathbf{T}\left(\mathcal{S}^C\right)|\mathbf{y}, V\right)$$

$$= I\left(\mathbf{T}\left(\mathcal{S}\right); \mathbf{y}|V\right) + I\left(\mathbf{T}\left(\mathcal{S}\right); \mathbf{T}\left(\mathcal{S}^C\right)|\mathbf{y}, V\right)$$

$$\leq I\left(\mathbf{x}\left(\mathcal{S}\right); \mathbf{y}|V\right) + I\left(\mathbf{x}\left(\mathcal{S}\right); \mathbf{x}\left(\mathcal{S}^C\right)|\mathbf{y}, V\right) \tag{59}$$

$$= I\left(\mathbf{x}\left(\mathcal{S}\right); \mathbf{y}|V\right) - I\left(\mathbf{x}\left(\mathcal{S}\right); \mathbf{x}\left(\mathcal{S}^C\right)|V\right) + I\left(\mathbf{x}\left(\mathcal{S}\right); \mathbf{x}\left(\mathcal{S}^C\right)|\mathbf{y}, V\right) \tag{60}$$

$$= I\left(\mathbf{x}\left(\mathcal{S}\right); \mathbf{y}|\mathbf{x}\left(\mathcal{S}^C\right), V\right). \tag{61}$$

We also used the fact that

$$I\left(A; B|C, D\right) = I\left(A; B|D\right) - I\left(A; C|D\right) + I\left(A; C|B, D\right) \tag{62}$$

and $I\left(\mathbf{T}\left(\mathcal{S}\right); \mathbf{T}\left(\mathcal{S}^C\right)|V\right) = 0$ since $\mathbf{T}\left(\mathcal{S}\right)$ and $\mathbf{T}\left(\mathcal{S}^C\right)$ are independent. Eqn. (59) follows from the data processing inequality. We proved that $I\left(\mathbf{T}\left(\mathcal{S}\right); \mathbf{T}\left(\mathcal{S}^C\right)|\mathbf{y}, V\right) = I\left(\mathbf{x}\left(\mathcal{S}\right); \mathbf{x}\left(\mathcal{S}^C\right)|\mathbf{y}, V\right)$ by Lemma 2.

Therefore $I\left(\mathbf{T}\left(\mathcal{S}\right); \mathbf{y}|\mathbf{T}\left(\mathcal{S}^C\right), V\right) \leq I\left(\mathbf{x}\left(\mathcal{S}\right); \mathbf{y}|\mathbf{x}\left(\mathcal{S}^C\right), V\right)$ and the theorem follows.

*Converse Proof*: We now prove the converse theorem.

Let $W^{(k)} \in \left\{1, \ldots, 2^{NR^{(k)}}\right\}$ be a random variable, where $k \in \{1, \ldots, K\}$. Then,

$$NR\left(\mathcal{S}\right) = H\left(W\left(\mathcal{S}\right)\right)$$

$$= H\left(W\left(\mathcal{S}\right)|\mathbf{y}_1^N, V_1^N\right) + I\left(W\left(\mathcal{S}\right); \mathbf{y}_1^N, V_1^N\right)$$

$$\leq H\left(W^{(1)}, \ldots, W^{(K)}|\mathbf{y}_1^N, V_1^N\right) + I\left(W\left(\mathcal{S}\right); \mathbf{y}_1^N, V_1^N\right)$$

$$\leq 1 + NRP_e^{(N)} + I\left(W\left(\mathcal{S}\right); \mathbf{y}_1^N, V_1^N\right) \tag{63}$$

$$= 1 + NRP_e^{(N)} + I\left(W\left(\mathcal{S}\right); \mathbf{y}_1^N|V_1^N\right), \tag{64}$$

where $R = \sum_{k=1}^{K} R^{(k)}$ and $W(\mathcal{S}) = \{W^{(k_1)}, \ldots, W^{(k_{|\mathcal{S}|})}\}$. Eqn. (63) follows from Fano's inequality and (64) follows from the independence between $W(\mathcal{S})$ and $V_n$.

$I\left(W(\mathcal{S}); \mathbf{y}_1^N | V_1^N\right)$ is given by

$$I(W(\mathcal{S}); \mathbf{y}_1^N | V_1^N) = \sum_{n=1}^{N} I\left(W(\mathcal{S}); \mathbf{y}_n | V_1^N, \mathbf{y}_1^{n-1}\right) \quad (65)$$

$$\leq \sum_{n=1}^{N} I\left(\mathbf{T}_n(\mathcal{S}); \mathbf{y}_n | V_1^N, \mathbf{y}_1^{n-1}\right) \quad (66)$$

$$= \sum_{n=1}^{N} \left[h\left(\mathbf{T}_n(\mathcal{S}) | V_1^N, \mathbf{y}_1^{n-1}\right) - h\left(\mathbf{T}_n(\mathcal{S}) | V_1^N, \mathbf{y}_1^n\right)\right]$$

$$= \sum_{n=1}^{N} \left[h\left(\mathbf{T}_n(\mathcal{S}) | V_n\right) - h\left(\mathbf{T}_n(\mathcal{S}) | V_n, \mathbf{y}_n\right)\right] \quad (67)$$

$$\leq \sum_{n=1}^{N} \left[h\left(\mathbf{T}_n(\mathcal{S}) | V_n, \mathbf{T}_n(\mathcal{S}^C)\right) - h\left(\mathbf{T}_n(\mathcal{S}) | \mathbf{y}_n, V_n, \mathbf{T}_n(\mathcal{S}^C)\right)\right] \quad (68)$$

$$= \sum_{n=1}^{N} I\left(\mathbf{T}_n(\mathcal{S}); \mathbf{y}_n | V_n, \mathbf{T}_n(\mathcal{S}^C)\right).$$

Eqn. (65) follows from the chain rule, and (66) follows from the data processing inequality. Eqn. (67) follows from the fact that $\mathbf{T}_n(\mathcal{S})$ is independent of the other variables except for $\mathbf{y}_n$, and (68) follows from the fact that the condition does not increase entropy and $\mathbf{T}_n(\mathcal{S})$ and $\mathbf{T}_n(\mathcal{S}^C)$ are independent.

We showed that $I\left(\mathbf{T}_n(\mathcal{S}); \mathbf{y}_n | \mathbf{T}_n(\mathcal{S}^C), V_n\right) \leq I\left(\mathbf{x}_n(\mathcal{S}); \mathbf{y} | \mathbf{x}_n(\mathcal{S}^C), V_n\right)$, and therefore,

$$R(\mathcal{S}) \leq \frac{1}{N} + RP_e^{(N)} + C(\mathcal{S}). \quad (69)$$

If $P_e^{(N)} \to 0$ as $N \to \infty$, then $R(\mathcal{S}) \leq C(\mathcal{S})$.

Now we prove (5). By [28], the capacity of any MAC can be written as the convex closure of the union of rate regions corresponding to the input product distribution satisfying the individual power constraint. Therefore to obtain the capacity region of our MIMO MAC, we can unionize all capacity polyhedrons about all covariance matrices satisfying the power constraint and take the convex closure. For a fixed set of covariance matrices, the capacity polyhedron is given by (4). The capacity region must be closed by Theorem 3 in [32]. We can show that this fixed capacity polyhedron is closed and that the union of these polyhedrons is also closed. This completes the proof. ∎

## Appendix B: Lloyd's Algorithm

We give Lloyd's algorithm to find the codebook that maximizes the minimum distance. The Grassmannian line packing problem approximately solved using Lloyd's algorithm for single user MIMO beamforming systems has been studied in [33, 34]. Based on the *Fubini-Study distance*, we define the distortion measure

$$d_1(\mathbf{G}, \mathbf{V}_q) = \arccos(|\det(\mathbf{G}^*\mathbf{V}_q)|), \tag{70}$$

where

$$\mathbf{G} = \begin{bmatrix} \mathbf{g}^{(1)} & \mathbf{0}_{M_t} & \cdots & \mathbf{0}_{M_t} \\ \mathbf{0}_{M_t} & \mathbf{g}^{(2)} & \cdots & \mathbf{0}_{M_t} \\ \vdots & \vdots & \ddots & \vdots \\ \mathbf{0}_{M_t} & \mathbf{0}_{M_t} & \cdots & \mathbf{g}^{(K)} \end{bmatrix}, \tag{71}$$

and each $\mathbf{g}^{(k)} \in \mathbb{C}^{M_t}$ is a unit-norm vector. To find an optimal codebook, we should solve the following optimization problem

$$\mathcal{V} = \underset{\mathcal{V}'}{\arg\min} \, E_{\mathbf{G}} [d_1(\mathbf{G}, \mathbf{V}(\mathbf{G}))], \tag{72}$$

where $\mathbf{V}(\mathbf{G}) = \underset{1 \leq q \leq 2^B}{\arg\min} \, d_1(\mathbf{G}, \mathbf{V}_q)$, $\mathbf{V}_q \in \mathcal{V}'$, and $\mathcal{V}'$ is a codebook of cardinality $2^B$. The optimization problem of (72) is difficult to solve because of the arccos function. Instead we define a modified distortion measure

$$d_2(\mathbf{G}, \mathbf{V}_q) = -\prod_{k=1}^{K} \left(\mathbf{v}_q^{(k)}\right)^* \mathbf{g}^{(k)} \left(\mathbf{g}^{(k)}\right)^* \mathbf{v}_q^{(k)}, \tag{73}$$

which is mathematically tractable to solve analytically. This is because (72) becomes

$$\begin{aligned}
\mathcal{V} &= \underset{\mathcal{V}'}{\arg\min} \, E_{\mathbf{G}} [\arccos(|\det(\mathbf{G}^*\mathbf{V}(\mathbf{G}))|)] \\
&\doteq \underset{\mathcal{V}'}{\arg\min} \, E_{\mathbf{G}} [-|\det(\mathbf{G}^*\mathbf{V}(\mathbf{G}))|] \tag{74} \\
&\doteq \underset{\mathcal{V}'}{\arg\min} \, E_{\mathbf{G}} [-|\det(\mathbf{G}^*\mathbf{V}(\mathbf{G}))|^2] \\
&= \underset{\mathcal{V}'}{\arg\min} \, E_{\mathbf{G}} \left[-\prod_{k=1}^{K} \left(\mathbf{v}^{(k)}(\mathbf{g}^{(k)})\right)^* \mathbf{g}^{(k)} \left(\mathbf{g}^{(k)}\right)^* \mathbf{v}^{(k)}(\mathbf{g}^{(k)})\right], \tag{75}
\end{aligned}$$

where $\mathbf{v}^{(k)}(\mathbf{g}^{(k)})$ is the $k$-th column of $\mathbf{V}(\mathbf{G})$. (74) follows from the fact that $0 \leq |\det(\mathbf{G}^*\mathbf{V}(\mathbf{G}))| \leq 1$ and $\arccos(\cdot)$ is a monotonically decreasing function on $[0, 1]$. Optimization of (75) is also easier than the optimization of (72) to obtain the analytical solution.

To minimize the average distortion, we apply Lloyd's algorithm. An intermediate step in applying Lloyd's algorithm is to solve the optimization problem

$$\begin{aligned}
\mathcal{V} &= \underset{\mathcal{V}'}{\arg\min} \, E_{\mathbf{G}} [d_2(\mathbf{G}, \mathbf{V})] \\
&= \underset{\mathcal{V}'}{\arg\min} \, E_{\mathbf{G}} \left[-\prod_{k=1}^{K} \left(\mathbf{v}^{(k)}\right)^* \mathbf{g}^{(k)} \left(\mathbf{g}^{(k)}\right)^* \mathbf{v}^{(k)}\right] \\
&= \underset{\mathcal{V}'}{\arg\min} \sum_{q=1}^{2^B} \left(-\prod_{k=1}^{K} \left(\mathbf{v}_q^{(k)}\right)^* E\left[\mathbf{g}^{(k)} \left(\mathbf{g}^{(k)}\right)^* \Big| \mathbf{G} \in \mathcal{G}_q\right] \mathbf{v}_q^{(k)}\right) P(\mathbf{G} \in \mathcal{G}_q) \\
&= \underset{\mathcal{V}'}{\arg\min} \sum_{q=1}^{2^B} \left(-\prod_{k=1}^{K} (\mathbf{v}_q^{(k)})^* \mathbf{R}_q^{(k)} \mathbf{v}_q^{(k)}\right) P(\mathbf{G} \in \mathcal{G}_q), \tag{76}
\end{aligned}$$

where $\mathcal{G} = \{\mathcal{G}_1, \cdots, \mathcal{G}_{2^B}\}$ is a specific Voronoi regions and $\boldsymbol{R}_q^{(k)} = E\left[\mathbf{g}^{(k)}\left(\mathbf{g}^{(k)}\right)^* | \mathbf{G} \in \mathcal{G}_q\right]$. To minimize the distortion measure, $\boldsymbol{v}_q^{(k)}$ can be chosen by the eigenvector associated with the maximum eigenvalue of $\boldsymbol{R}_q^{(k)}$. Therefore we can obtain $\mathcal{V} = \{\boldsymbol{V}_1, \ldots, \boldsymbol{V}_{2^B}\}$.

In general, the average distortion converges, but the minimum distance $\delta_{FS}$ does not converge as iteration proceeds. $\delta_{FS}$ increases during the first few step, and has tendency to increase as iteration proceeds but may oscillate slightly after some iteration [33]. In this case, as indicated in [33], we save all intermediate minimum distances during the iterations and at some fixed iteration number, we obtain the maximum of the minimum distances from the previous saved minimum distances. We also obtain the codebook that corresponds to the index that maximizes the minimum distance.

[22] W. Santipach and M. L. Honig, "Asymptotic performance of MIMO wireless channels with limited feedback," in *Proc. IEEE Mil. Comm. Conf.*, vol. 1, Oct. 2003, pp. 141–146.

[23] ——, "Asymptotic capacity of beamforming with limited feedback," in *Proc. IEEE Int. Symp. Info. Th.*, Jun. 27-Jul. 2 2004, p. 289.

[24] ——, "Achievable rates for MIMO fading channels with limited feedback and linear receivers," in *Proc. Int. Symp. Spread Spectrum Techniques and Applications*, Aug. 30-Sept. 2 2004, pp. 1–6.

[25] C. K. Au-Yeung and D. J. Love, "On the performance of random vector quantization limited feedback beamforming in a MISO system," *IEEE Trans. Wireless Comm.*, accepted for publication.

[26] G. J. Foschini and M. J. Gans, "On limits of wireless communications in a fading environment when using multiple antennas," *Wireless Personal Communications 6*, pp. 311–335, 1998.

[27] D. Shiu, G. J. Foschini, M. J. Gans, and J. M. Kahn, "Fading correlation and its effect on the capacity of multielement antenna systems," *IEEE Trans. Commun.*, vol. 48, pp. 502–513, Mar. 2000.

[28] A. Goldsmith, S. Jafar, N. Jindal, and S. Vishwanath, "Capacity limits of MIMO channels," *IEEE Jour. Select. Areas in Commun.*, vol. 21, no. 5, pp. 684–702, June 2003.

[29] D. J. Love and R. W. Heath Jr., "Limited feedback diversity techniques for correlated channels," *IEEE Trans. on Veh. Technol.*, vol. 55, no. 2, pp. 718–722, Mar. 2006.

[30] N. Jindal and A. Goldsmith, "Dirty-paper coding versus TDMA for MIMO broadcast channels," *IEEE Trans. Info. Th.*, vol. 51, no. 5, pp. 1783–1794, May 2005.

[31] T. Cover and J. Thomas, *Elements of Information Theory*. New York, NY: Wiley, 1991.

[32] S. Verdú, "Multiple-access channels with memory with and without frame synchronism," *IEEE Trans. Info. Th.*, vol. 35, no. 3, pp. 605–619, May 1989.

[33] P. Xia, S. Zhou, and G. B. Giannakis, "Achieving the Welch bound with difference sets," *IEEE Trans. Info. Th.*, vol. 51, no. 5, pp. 1900–1907, May 2005.

[34] P. Xia and G. B. Giannakis, "Design and analysis of trasmit-beamforming based on limited-rate feedback," *IEEE Trans. Sig. Proc.*, vol. 54, no. 5, pp. 1853–1863, May 2006.
20

Figure 1: MIMO MAC model with CSIT and CSIR.

Figure 2: Channel model of the communication system with CSIT and CSIR.

Figure 3: Equivalent channel model of the communication system with CSIT and CSIR.

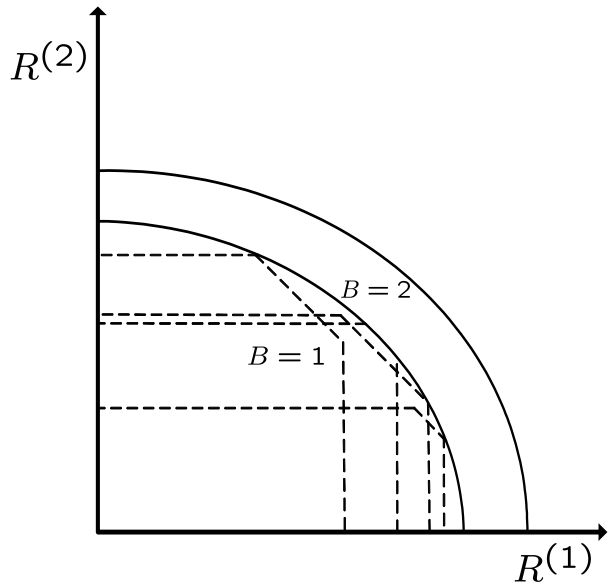

Figure 4: Capacity region for MIMO MAC according to the cardinality of $\mathcal{U}$.

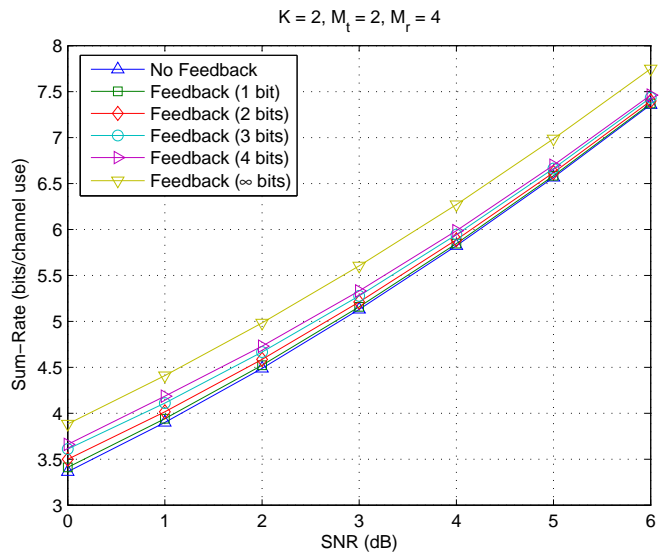

Figure 5: Sum-rate capacity of a (2,2,4) system using a covariance codebook under a sum power constraint.

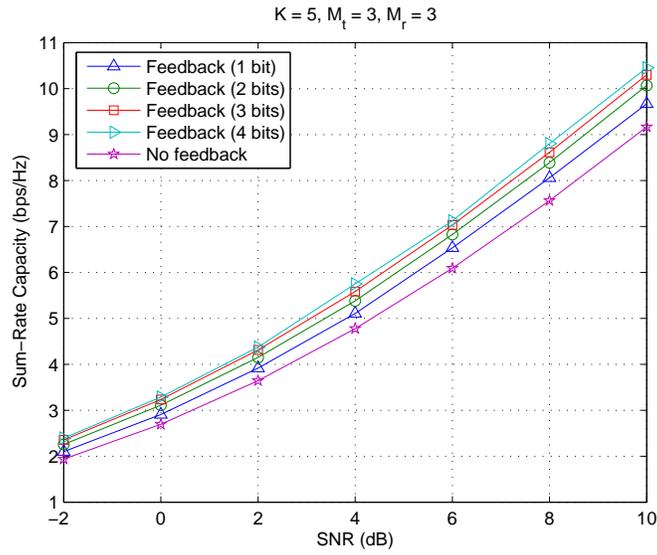

Figure 6: Sum-rate capacity of a (5,3,3) system using an eigenbeamforming codebook under a sum power constraint.

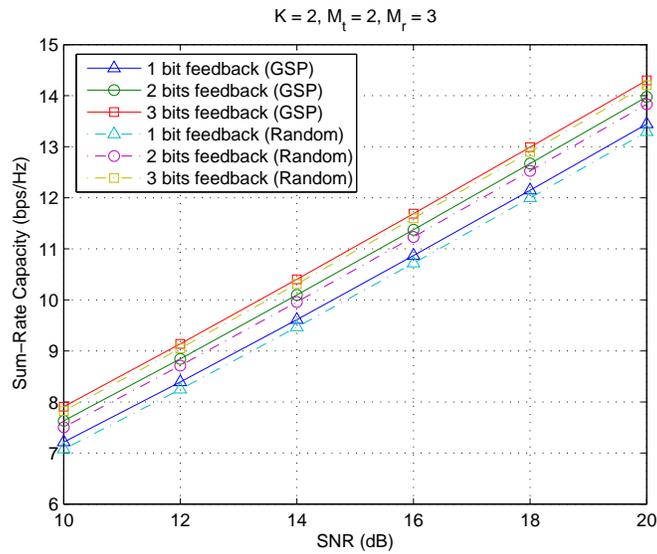

Figure 7: Sum-rate capacity of a (2,2,3) system using a Grassmannian beamforming codebook and random beamforming codebook with random power allocation.

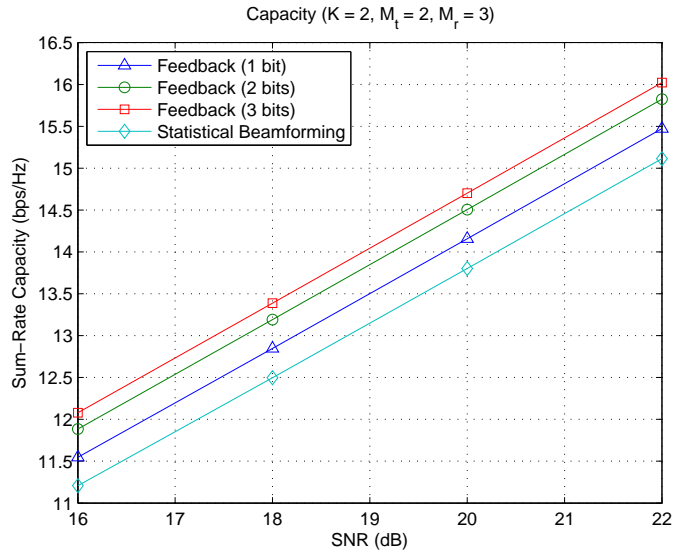

Figure 8: Sum-rate capacity of (2,2,3) systems using Grassmannian beamforming and statistical beamforming.

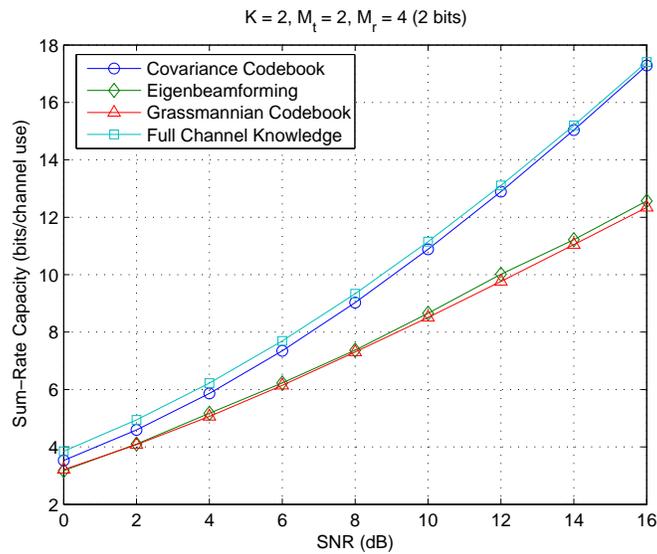

Figure 9: Sum-rate of (2,2,4) systems for the covariance codebook, the eigenbeamforming codebook, and the Grassmannian beamforming codebook.

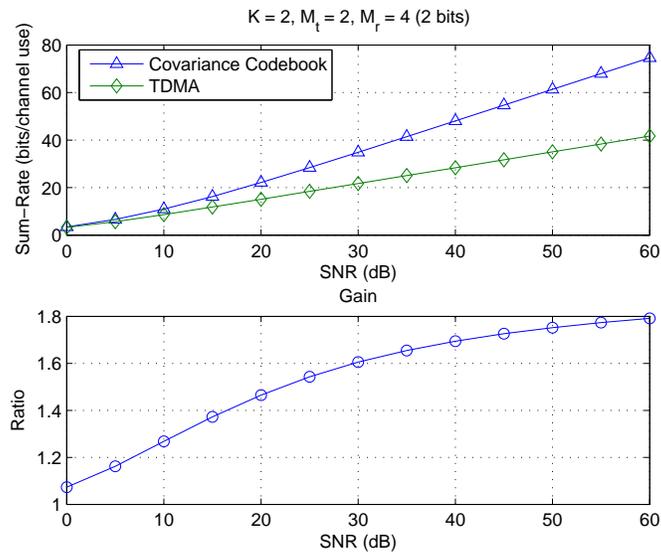

Figure 10: Sum-rate of (2,2,4) systems for the covariance codebook and TDMA scheme, and their comparison.